\documentclass[preprint]{aastex}

\usepackage[]{natbib}
\usepackage{epsfig}
\usepackage{graphics}
\usepackage{amssymb}

\newcommand{\ie}{{\rm i.e.}}
\newcommand{\Mo}{\ensuremath{M_\odot}}

\newcommand{\Lo}{\ensuremath{L_\odot}}

\newcommand{\asec}{\ensuremath{\arcsec}}
\newcommand{\ml}{\ensuremath{M/L}}
\newcommand{\mlb}{\ensuremath{M/L_{B}}}

\newcommand{\mlo}{\ensuremath{\Mo/\Lo}}
\newcommand{\hmlo}{~\ensuremath{{h}{\mlo}}}

\newcommand{\hkpc}{~\ensuremath{{h}^{-1}{\rm kpc}}}

\newcommand{\omegamnought}{\ensuremath{\Omega_{m0}}}

\newcommand{\scrit}{\ensuremath{\Sigma_{\rm crit}}}
\newcommand{\scritrat}{\ensuremath{\Sigma/\Sigma_{\rm crit}}}

\def \figwidth {\linewidth}

\shorttitle{Title}
\shortauthors{Wilson \etal}

\begin{document}

\title{MASS AND LIGHT IN THE UNIVERSE\altaffilmark{1}}

\author{Gillian Wilson\altaffilmark{2}}

\altaffiltext{1}{This article is based on the third (Wilson, Kaiser, \& Luppino, 2001, Ap.J., 556, 601) in a series of papers describing results from an ongoing project whose principle aim is to investigate the cosmic shear pattern caused by gravitational lensing from the large-scale structure of the Universe.
}

\altaffiltext{2}{Physics Department, Brown University, 182 Hope Street, Providence, RI 02912 : gillian@het.brown.edu}

\begin{abstract}
We present a weak lensing and photometric study of 
six $0\fdg5 \times 0\fdg5$
fields observed at the 3.6m CFHT telescope using the $8192 \times 8192$
pixel UH8K CCD mosaic camera. The six fields were chosen to be ``blank fields'' \ie\ 
representative views of the
Universe. The fields were observed for a total of 2 hours each in $I$ and $V$,
resulting in catalogs containing $\sim 20$,$000$ galaxies per passband per field.
We used $V-I$ color and $I$ magnitude to select bright early type galaxies 
at redshifts $0.1 < z < 0.9$. 
We measured the gravitational shear from faint galaxies in the range $21 < m_{I} < 25$
from a composite catalog and found a strong correlation with that predicted from the early types
if they trace the mass with $\mlb \simeq 300\pm75 \hmlo$ for a flat ($\Omega_{{\rm m}0} = 0.3, \Omega_{\lambda 0} = 0.7$) lambda cosmology and
$\mlb \simeq 400\pm100 \hmlo$ for Einstein-de Sitter.
We made two-dimensional reconstructions of 
the mass surface density. Cross-correlation of the measured 
mass surface density with that predicted from the
early type galaxy distribution showed a strong peak at zero lag (significant at the $5.2\sigma$
level). We azimuthally averaged the cross- and auto-correlation functions. 
 We concluded that the
profiles were consistent with early type galaxies tracing  mass 
on scales of
$\geq45\asec$ ($\geq200 \hkpc$ at $z=0.5$). 
We sub-divided our bright early type galaxies by redshift and obtained similar conclusions.
These $\mlb$ ratios imply $\omegamnought \simeq 0.10\pm0.02$ ($\omegamnought \simeq 0.13\pm0.03$ for Einstein-de Sitter) of closure density.

\end{abstract}

\section{Mass Surface Density Predictions from Luminous Matter and from 
 Lensing}

\begin{figure}
\centering\epsfig{file=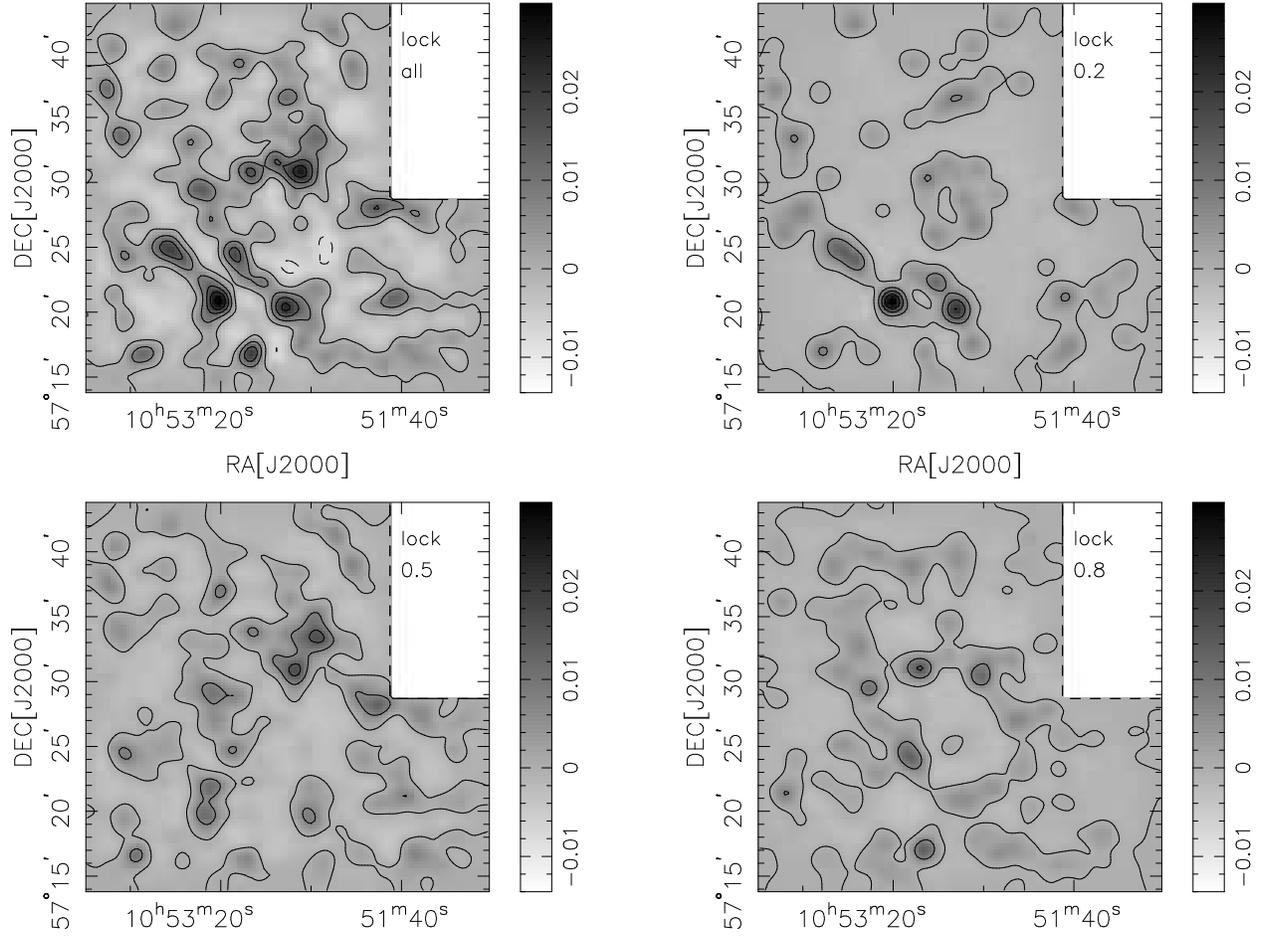,width=\figwidth}
\caption[f1.ps]{
 Upper left panel shows the predicted mass surface density using early type
galaxies selected by $V-I$ color, and $\mlb = 300\hmlo$ for the Lockman 
field (one of our six pointings).  
The image has been smoothed with a $45\asec$ Gaussian filter.
The mean has been subtracted from the image. The wedge shows the 
calibration of the grayscale
and the contour separation is $0.007\times\scritrat$.
The remaining three panels show the predicted surface mass density using early type
galaxies and $\mlb = 300 \hmlo$ but subdividing the galaxies into 
$z = 0.2, 0.5, 0.8 \pm0.15$.

\label{fig:pred_lock}
}
\end{figure}

Our analysis differed from other approaches in that we used 
$V-I$ color and $I$ magnitude to reliably 
select bright early type galaxies 
at redshifts $0.1 < z < 0.9$. We generated predictions of the dimensional 
mass surface density $\kappa$ (where $\kappa = \scritrat$,
the physical mass per unit area in units of the critical surface density)
from $I$-band galaxy luminosity, assuming a constant $\mlb$. The implicit assumption is that 
optical  early type galaxy luminosity is an unbiased tracer of the mass. We adopted a redshift
distribution for the faint source galaxies (needed to calculate $\scrit$) 
based on spectroscopic redshifts from Len Cowie's (ongoing) 
Hawaii galaxy survey.
The upper left panel of Fig.~\ref{fig:pred_lock} 
shows the resulting
surface mass density $\kappa$-from-light ``map'' 
for one of the six fields. 


\begin{figure}
\centering\epsfig{file=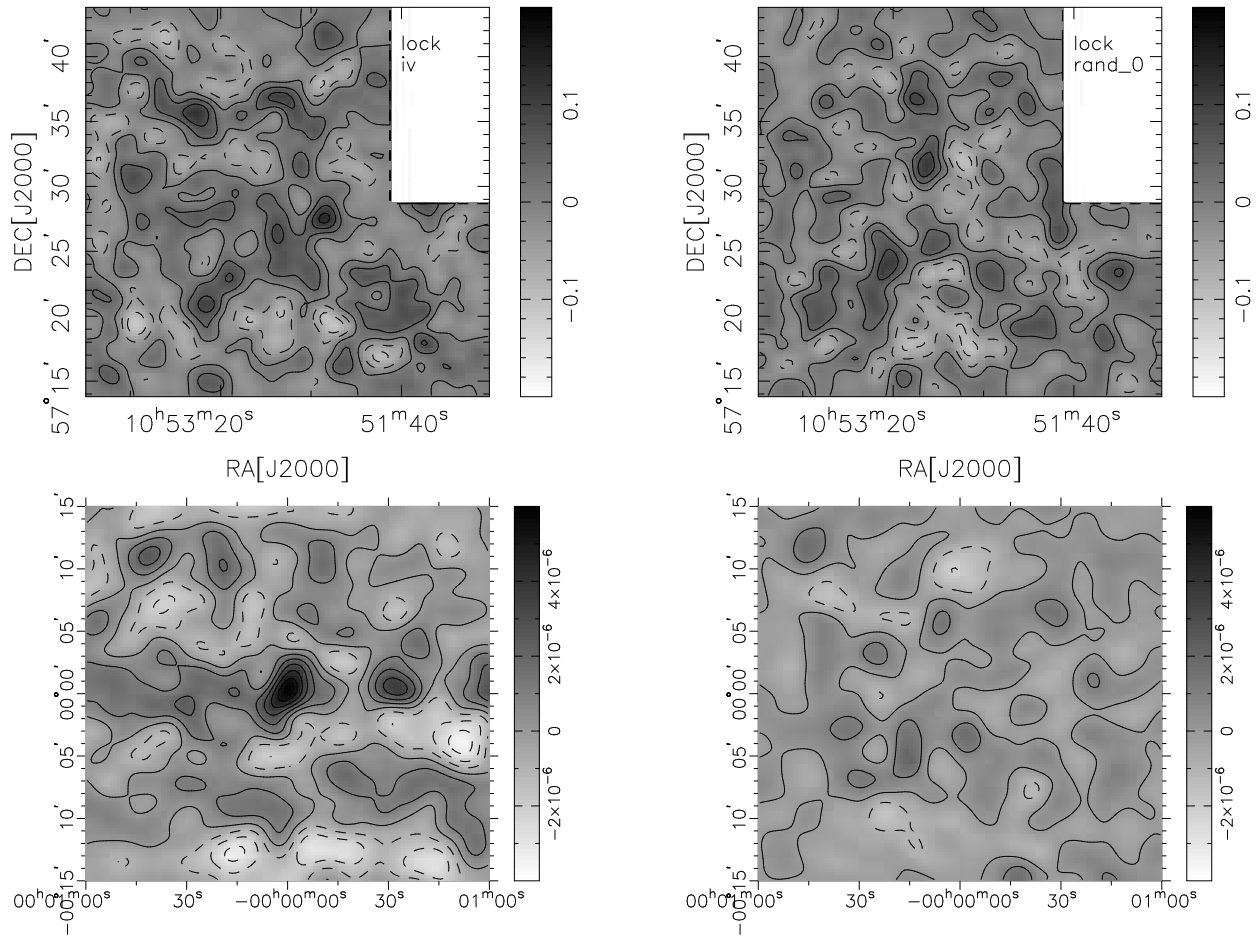,width=\figwidth}
\caption[f2.ps]{
Upper panels show reconstructions of mass surface density inferred from lensing 
for the same (Lockman) field. The upper left panel shows the
reconstruction from a composite $IV$ catalog, and the upper right panel shows the
reconstruction from the same catalog with randomized ellipticities, indicating the expected
noise fluctuations due to intrinsic random galaxy shapes. The reconstructions have been smoothed
with a $45\asec$ Gaussian filter.
\vspace{2mm}
\newline
The cross-correlation of light (Fig. 1 upper left) with mass reconstruction (This Fig. upper left) averaged over all six
pointings is shown in the
lower left panel. The peak at zero lag is significant at the $5.2\sigma$ level.
The lower right panel shows the lack of correlation, as expected, when 
the randomized catalog reconstruction (This Fig. upper right) is used instead.

\label{fig:poster1}
}
\end{figure}

From galaxy shear estimates we also constructed 
2D mass surface density reconstructions \emph{directly}.
The upper left panel of Fig.~\ref{fig:poster1} shows the 
surface density map inferred from lensing
for the same field. For comparison,
the upper right panel shows the typical (high) level of noise.

In view of the difficulty of measuring mass directly from the maps, and 
since we have a large area containing many structures, we attempted
to better reveal the signal by cross-correlating light and mass.
We therefore cross-correlated the luminosity associated 
 with foreground galaxies with the mass inferred from the background 
galaxy shear estimates. Our aim was firstly to test the hypothesis of a constant
mass-to-light ratio which is independent of scale, and also to determine the value of $\mlb$,
the constant of proportionality between mass and light.

\section{Mass-Light Cross-Correlation, Inferred Mass Density \omegamnought, and Conclusions} 

 Cross-correlation of the measured 
mass surface density with that predicted from the
early type galaxy distribution showed a strong peak at zero lag (significant at the $5.2\sigma$
level) as shown in the lower left panel of Fig.~\ref{fig:poster1}.
 That early type galaxy luminosity and total mass show such a strong correlation is our central result.
The correlation strength at zero lag implies a 
$\mlb \simeq 300\pm75 \hmlo$ for a flat ($\Omega_{{\rm m}0} = 0.3, \Omega_{\lambda 0} = 0.7$) lambda cosmology and $\mlb \simeq 400\pm100 \hmlo$ for Einstein-de Sitter.

In order to determine if $\ml$ ratio varies with scale, we then
azimuthally averaged the cross- and auto-correlation functions. 
The cross- and auto- correlation functions had very similar
profiles. We concluded that the
profiles are consistent with early type galaxies tracing  mass 
on scales of
$\geq45\asec$ ($\geq200 \hkpc$ at $z=0.5$). 
We sub-divided our bright early type galaxies by redshift (same as Fig.~\ref{fig:pred_lock}) and obtained similar conclusions.


We then calculated $\omegamnought$ using
$\omegamnought =  \rho_{E}/ \rho_{\rm crit}$
and
$\rho_{E} =  (M/L)_{E}{\mathcal L}_{E} = (M/L)_{E} \phi_{\star E} L_{\star E} \Gamma(\alpha_{E}+2)$
where ${\mathcal L}_{E}$ is the measured $B$-band luminosity density of
the universe for early type galaxies. We concluded that
$\omegamnought \simeq 0.10\pm0.02$ ($\omegamnought \simeq 0.13\pm0.03$ for Einstein-de Sitter) of closure density.

The global density parameter we obtain appears low compared to other estimates. 
This is not because our early type $\mlb$ value is low  
 but because we
do not assign the same $\mlb$ to late types as to early types. In the scenario we propose, late types are 
assumed to have very similar luminosities to early types. The difference is that late types have much less 
mass associated with them and hence their $\mlb$ ratio is much lower. The 
analysis in this paper
assumes that they have negligible $\mlb$ compared to early types.


In summary, we found that the majority of mass in the Universe is associated with 
early 
type galaxies. On scales 
of $\geq 200\hkpc$ (the smoothing scale at $z=0.5$) it appears that their light traces the underlying 
mass distribution with a constant $\mlb =300 -400 \pm 100 \hmlo$, depending on
cosmology.
As with several other recent results our data argues against an $\omegamnought=1$ universe.

\end{document}